# Experimental Results for the Sensitivity of a Low Noise Aperture Array Tile for the SKA

E. E. M. Woestenburg, L. Bakker, and M. V. Ivashina

*Abstract*—Aperture arrays have been studied extensively for application in the next generation of large radio telescopes for astronomy, requiring extremely low noise performance. Prototype array systems need to demonstrate the low noise potential of aperture array technology. This paper presents noise measurements for an Aperture Array tile of 144 dual-polarized tapered slot antenna (TSA) elements, originally built and characterized for use as a Phased Array Feed for application in an L-band radio astronomical receiving system. The system noise budget is given and the dependency of the measured noise temperatures on the beam steering is discussed. A comparison is made of the measurement results with simulations of the noise behavior using a system noise model. This model includes the effect of receiver noise coupling, resulting from a changing active reflection coefficient and array noise contribution as a function of beam steering. Measurement results clearly demonstrate the validity of the model and thus the concept of active reflection coefficient for the calculation of effective system noise temperatures. The presented array noise temperatures, with a best measured value of 45 K, are state-of-the-art for room temperature aperture arrays in the 1 GHz range and illustrate their low noise potential.

## I. INTRODUCTION

FOR the Square Kilometre Array (SKA, [1]), the next generation of large radio telescopes with two orders of magnitude increase in sensitivity over existing telescopes, the radio astronomical community is considering the use of dense Aperture Arrays (AAs) for the SKA mid-frequency range from 400 MHz to 1400 MHz. Considerable effort has been put in the development of aperture arrays over the last ten years. Low frequency aperture arrays up to a few 100 MHz have been developed and are already in operation ([2], [3]) and several prototypes for higher frequencies up to 1.5 GHz have been built ([4], [5]), while development for dense aperture arrays over the frequency range from 100 MHz to 1500 MHz is continuing ([6], [7]). The arrays for the SKA-mid frequency range will consist of a large number of 1 m² tiles with approximately 100 flat antennas per tile with LNAs and receivers, of which the signals will be combined to form a radio telescope with a (collecting) area of 500 m², as an alternative to reflector telescopes. Advantages of this AA-concept over conventional reflector antennas are the wide field of view, the possibility to avoid mechanical steering and maintenance (as beams will be formed and steered electronically) and the opportunity to observe with a large number of beams simultaneously. On the other hand, the arrays will have to operate at ambient temperature, because cooling cannot be realized at reasonable cost and complexity, due to the large number of antennas and LNAs. At the same time the array sensitivity is of utmost importance and is determined by the ratio of effective collecting area and system temperature Aeff/Tsys. The large collecting area of one square kilometer is the main reason for a greatly enhanced sensitivity, but requires that the system noise temperature will be at a competitive level with that of conventional radio telescopes.

Emphasis in the development of AA-tiles for the concept demonstrator systems has thus far not been on achieving the lowest possible noise temperatures, which is a challenge because of the room temperature operation. Reported noise temperatures until now for aperture array tiles and systems in the 1 GHz frequency range are relatively high, around 170 K in [8] and approximately 100 K in [9], compared to the ultimately required 40 K maximum system noise temperature for the SKA near 1 GHz. Nevertheless, the results in [9] are similar to predicted values in [10] and a decreasing trend in noise temperature is obvious ([11]).

The focus for the AA-tiles and systems in [4] and [5] has been on proof of principle, large scale systems and adequate production techniques, limited costs and operability. The latter have been demonstrated in [9]. At the same time development of low noise AAs has been progressing at different groups within the SKA community. A similar development has been ongoing for Phased Array Feed (PAF) systems for reflector telescopes ([12]-[15]), while wide field imaging has been demonstrated with PAF prototype systems on such telescopes ([15]-[17]). Simultaneous with the development and construction of AA- and PAF-demonstrator systems progress was made in the theoretical analysis and modeling of the noise properties of phased array systems with high sensitivity, in particular with respect to the effect of noise coupling between the antenna elements ([18]-[22]). The understanding, resulting from this theoretical work, has favored the realization of prototype arrays with a factor 2, respectively 4 improvement in noise performance with respect to [9] and [8]. This paper discusses state-of-the-art results of noise measurements with a prototype AA-tile, which enable verification of the noise models.

The AA results presented in this paper have profited from the development of a tile for a PAF system (APERTIF,

APERture Tile In Focus, [17]), which will replace the existing 21 cm single pixel feeds of the Westerbork Synthesis Radio Telescope (WSRT), and should have competitive performance with the present 21 cm cryogenic receiver system. This development resulted in room temperature receivers and LNAs for an APERTIF prototype tile, giving a measured system temperature in the telescope below 68 K for an on-axis beam at 1.4 GHz ([16], [17]), using LNAs with close to 40 K noise temperature ([23]).

The construction of the PAF tile with individual antennas and receiver chains, allowed its use as an AA-tile, while the measured results as a PAF gave rise to great expectations with respect to the noise performance as an AA-tile. A description of the prototype tile as AA-tile, as well as the noise measurement set-up will be given in section II. Measurement results of the array noise temperature will be presented in section III, showing state-of-the-art noise performance for a room temperature AA-tile. Comparable performance has been recently shown for a smaller array in [24], but only measured data for single channel receiver noise temperatures are presented there. Results from our measurements are full array noise temperatures and will be presented for various beam configurations, both with analog and digital beam forming. In section IV a noise temperature budget for the AA-tile will be presented and a comparison with simulated values is made. The latter are calculated using the equivalent system noise model in [22] and take into account the variation of system noise temperature as a function of scan angle ([20], [21]). The results presented here have a lowest measured value of 45 K at 1200 MHz and illustrate the low noise potential of AAs for the SKA.

## II. MEASUREMENT METHOD

The APERTIF tile used for the aperture array measurements consists of a total of 144 Vivaldi antenna elements and LNAs, configured in a two-dimensional array with 8x9 elements with 11 cm spacing, for each of the two linear polarizations ([25]). The antenna array operates in the frequency range 1000 – 1750 MHz with the optimal signal-to-noise performance around 1.4 GHz at which the element spacing is close to the half wavelength. Details of the system for use as a PAF can be found in [16-17] and [26]. For noise measurements as an aperture array, using the Y-factor hot/cold method ([8], [15]), the tile is placed horizontally on the ground as shown in Fig. 1. For the measurements two methods were followed, one using analog beam forming with 2x2 and 4x4 elements, the other using digital beam forming for various beam configurations. To assess the low noise potential of the tile, initial measurements were done with analog beam forming only. The output signals from the LNAs of a 2x2 and a 4x4 array in the centre of the tile were added with in-phase combiners, forming broadside beams. The analog output signals of the beam formers were fed to the input of an Agilent Noise Figure Meter 8970B and the noise temperature was determined with the Y-factor method, using the loads described next for the digital processing method. The first very promising measurement results urged the use of a more flexible system, with which beams in any direction could be formed. The digital beam forming and processing system of the APERTIF-prototype (see [17] for a description of the digital processing hardware and architecture) was subsequently made available for off-line data processing and calculation of the beam former output noise power (based on the measured spectral noise-wave correlation matrix of the array-receiver system according to the procedure in [26]). In order to compare to the previous measurements and verify the digital processing, the outputs of the analog beam formers were first connected to one channel of the digital processing system.

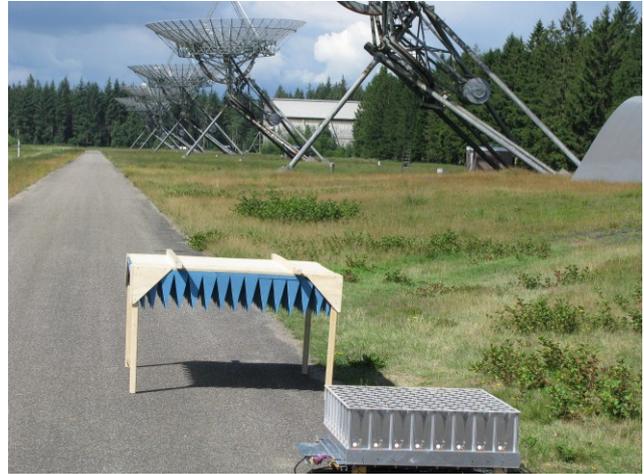

Fig. 1  Picture of the prototype array situated for hot/cold measurements, with the hot load behind the tile. In the background some of the WSRT telescopes are visible

For the digital processing method, a total of 49 individual antenna elements and LNAs (limited by the number of available receivers at the time of the measurements) are connected via 25 m long coaxial cables to a back-end. The electrical lengths of the cables have been made equal within 5º. All receiver channels, including the cables, were calibrated with respect to a reference channel. The differential phase stability of the receiver channels is 1º. The back-end electronics is located in a shielded cabin, with down converter modules and digital processing hardware ([17]). Data are taken with the array facing the (cold) sky as a 'cold' load, after which a room temperature absorber panel is placed over the array for the measurement with the 'hot' load. This 1.2 m x 1.2 m panel, shown behind the array in Fig. 1, is slightly oversized with respect to the outer dimensions of the tile (1 m x 1m), to avoid the edge elements seeing the cold sky beyond the edge of the absorber panel. Once placed over the array, the panel leaves approximately 0.3-0.5 wavelength (10 cm) space between the tips of the absorbers and the aperture plane of the antenna array. At this distance the absorbers do not influence the measured array S-parameters and provide a load with 30 dB return loss at normal incidence. Transmission through the absorbers, which have a depth of 18 cm and a 2.5 cm thick absorptive backing, is negligible at these low frequencies.

The data processing takes into account correlations between individual receiver channels in calculation of the beam former output noise power. This is done through performing the eigenvalue decomposition of the measured spectral noise-wave correlation matrix of the array-receiver system and taking its dominant eigenvector. This vector holds the noise-wave amplitudes at the receiver outputs that arise due to all internal and external noise sources and is a function of frequency. This vector is sometimes named the 'calibration vector' ([28]), as this procedure ensures that practical errors (such as electronic gain differences and non-equalization of the cables) do not impose limits on the accuracy of the noise temperature measurements and desired beam forming direction. Using off-line digital processing, beams with a combination of any of the 49 active elements can be formed and beams may be scanned in any direction by applying weights to the elements of this noise-wave vector (see Eq. 1 and 2 in [26]). In this way the array noise temperature as a function of frequency from 1.0 to 1.8 GHz has been determined for 2x2, 4x4, 5x5 and 7x7 element arrays, looking at broadside. Also the array noise temperatures as a function of scan angle for the 4x4 and 7x7 element arrays have been determined.

III. EXPERIMENTAL RESULTS

For all measurements a cold sky load temperature of 4 K was assumed, slightly above the cosmic background noise of 2.7 K. The 4 K cold load temperature is considered to be a realistic value, as long as the beam is not directed at the horizon or directly pointed at the sun, which may be distinguished as an approximately 200 K source for the 49-element array in Fig. 5. The ambient temperature was taken as the hot load temperature, approximately 300 K during the measurements. The uncertainty in the measured ambient temperature is smaller than 1 K. Another source of error in the calculation of the noise temperature is the accuracy of the power measurements and the resulting Y-factor. The accuracy of the Y-factor measurement is estimated at ±0.1 dB, which would result in a maximum error of ±2 K for the measured Y-factors around 8 dB and the given load temperatures. The total results in a maximum absolute error in the noise measurement of approximately 5 K. The error between measurements at different frequency points, performed in the same measurement cycle, is much smaller and is estimated at 1 K maximum.

It should be pointed out that the error in the hot load temperature is mitigated considerably by the large Y-factors measured. An error due to the measurement accuracy of the ambient temperature or the construction and size of the absorber panel would give a 1 K noise temperature error for a 6 K error in the hot load temperature, based on the measured Y-factors around 8 dB. A possible error due to the absorber panel would alter (lower) the hot load temperature, which could result in the presented noise temperatures being too pessimistic. In practice this effect appears to be negligible.

*A. Array noise temperatures at broadside*

Fig. 2 shows a comparison of the measurement results for the analog and digital beam forming, for the broadside beams formed using 4 and 16 active elements in the centre of the array. The passive elements were connected to 75 Ω loads during these initial measurements, which were performed mainly to verify the digital processing method. The measured array noise temperatures for the 4- and 16-element arrays are very similar, for both analog and digital beam formers, with some minor differences due to small differences between these beam formers and the shape of the beams in combination with environmental factors. The results are consistent with the data (not shown here) taken with the analog processing system described in the previous

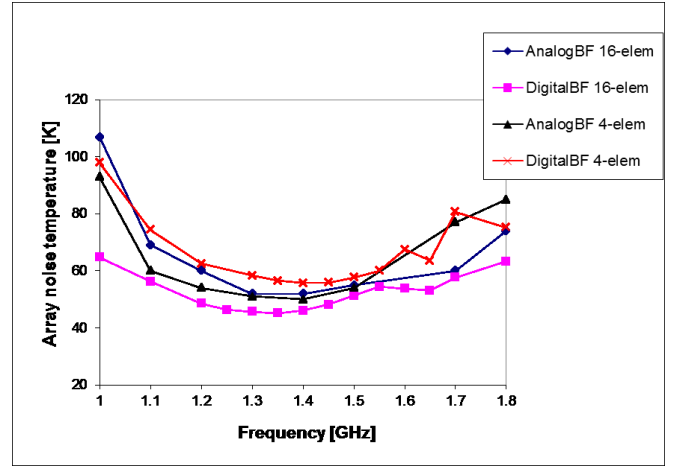

Fig. 2 Comparison of results for analog and digital beam forming for 4- and 16-element arrays

section and validate the results from the digital processing method within the 5 K error bars.

Using the digital processing method four different array beams have been formed with 4, 16, 25 and 49 active elements. During the measurements producing the digital data, all antenna elements were connected to their LNA and subsequent receiver chain, but for the unused elements the weights were set to zero during the data processing. Complex weights for the other, active, elements were set to direct the beam to broadside or any other desired direction.

Fig. 3 shows the measured noise temperatures for broadside beams with various numbers of active elements, compared to simulation results for the 49-element beam. The 25-element array shows the lowest noise on average as a function of frequency, close to that of the 16-element array, both being slightly better than the 49-element array. This may be explained by the particular noise coupling contribution for the 49-element array near broadside. At slightly different beam angles for the 49-element array this contribution is reduced and results in the same minimum value at e.g. 1.4 GHz as for the 16- and 25-element arrays.

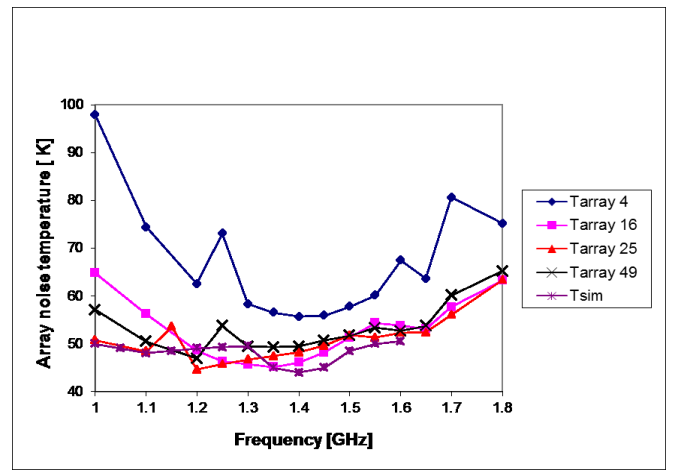

Fig. 3 AA noise temperature for broadside beams with various numbers of active elements, compared to simulation results for the 49-element beam

The 4-element array has a broader beam with 9-12 dB directivity at 1-1.6 GHz as compared to the larger arrays which have directivity values higher than 15 dB at the lowest frequency, and hence suffers from considerable noise contributions from the environment, even when it is directed at broadside. The simulated noise temperatures for the 49-element array, as well shown on Fig. 3, compare well within the absolute measurement accuracy of 5 K to the measured values for that array, as well as those for the 16- and 25-element arrays, over most of the frequency range.

### B. Array noise temperature as a function of scan angle

An important property of AAs is the varying noise coupling between the antenna elements as a function of scan angle, described by the active reflection coefficient ([20]-[22]). Using the array and LNA properties a maximum of 13 K is calculated for the 49-element array at 45° scan angle, 25 % of the system noise budget (see the simulation results in Fig. 4 and in section IV). It is therefore interesting to see how the measured array noise temperature varies with scan angle and beam steering in general. Fig. 4 shows the variation in noise temperature at 1400 MHz for the 16- and 49-element arrays, for a maximum scan angle (from broadside) of 85°. Up to scan angles of 40° the noise temperatures only slowly vary due to the changing noise coupling, quite well in agreement with the simulations. For larger scan angles the broad beams introduce noise from the environment, which dominates over the calculated variation in noise coupling. Obviously the narrower 49-element beam allows further scanning without increase in noise than the 16-element beam. It is also shown that the narrower beam more effectively samples the environmental temperature near the horizon, giving a higher system noise temperature.

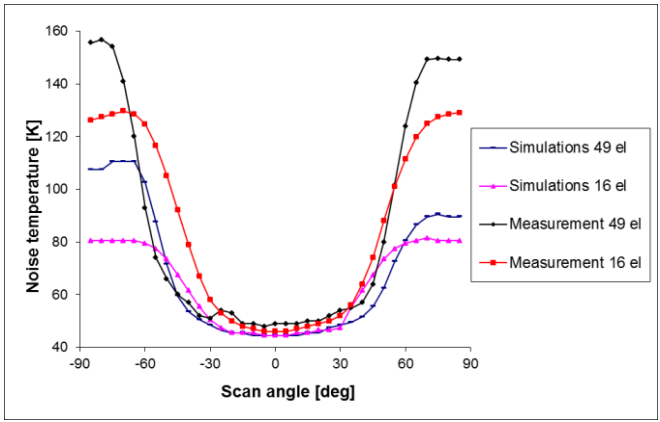

Fig. 4 Noise temperatures as a function of scan angle for 16- and 49-element arrays at 1400 MHz, compared to simulation results for both arrays. The simulations include noise coupling contributions and the noise pick-up from the environment at the horizon. The detailed effects of the location and height of trees and buildings are not taken into account in these simulations

### C. Results for two-dimensional scanning

In Fig. 5 a plot is shown of the system noise temperature at 1400 MHz for two-dimensional scanning of the 49-element array. Over most of the scanning range the noise temperature is well below 80 K, with lowest values under 50 K. As expected a strong increase in noise temperature is visible near the horizon, with extended hot regions at the location of one of the WSRT telescopes (top right) and the sun (lower right).

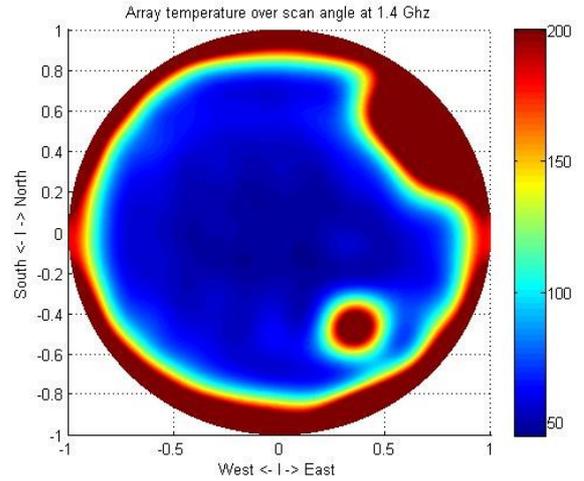

Fig. 5 Two-dimensional scan of Tsys at 1400 MHz for a 49-element array, showing the location of one of the WSRT telescopes (top right) and the sun

### IV. NOISE BUDGET AND COMPARISON WITH MODELING RESULTS

Table I shows the noise budget (partly based on simulations) of the prototype AA-tile at 1400 MHz, with the noise coupling contribution for a few scan angles. The equivalent system representation of [22] was used and the simulations were performed with the numerical approach in [26-27]. The simulations did not take into account the effect of the sky noise and contributions from obstacles (trees and telescopes/buildings) near the horizon. The array noise temperature was calculated as a sum of several noise contributions due to external and internal noise sources, using the equivalent system representation in [22]. The external noise contribution is the ground noise picked up due to antenna back radiation, which was computed from the simulated illumination pattern of the antenna array for the specified beam former weights. The internal noise contribution includes two components:

- the thermal antenna noise due to the losses in the conductor and dielectric materials of TSAs and microstrip feeds. The conductor losses are computed through the evaluation of the antenna radiation efficiency using the methodology detailed in [29] and the dielectric losses are computed based on the experimental evaluation of the feed loss [30].
- multi-channel receiver noise which is calculated using CAESAR software that is an array system simulator, developed at ASTRON ([27]). This noise component accounts for the antenna-LNA impedance noise mismatch effect and minimal noise of LNAs. This noise coupling component represents a combined effect of the noisy LNAs and active reflection coefficients of the array elements that are frequency and weighting dependent. It was computed from the noise-wave covariance matrix of the antenna-receiver system (see Eq.1 and fig.1 in [26]) that was defined in the absence of the external noise sources and antenna thermal noise.

The antenna noise $T_{ant\_noise}$ in Table I comprises contributions due to the ground noise pick up (1 K), losses in the conductor and dielectric material of the antenna and the microstrip feed. $T_{LNA}$ is the measured LNA noise temperature in a 50 Ω system, including the receiver second stage contribution. $T_{cpl\_simul}$ is the simulated noise coupling contribution. Adding these contributions leads to the simulated array noise temperature

$$T_{total\_simul} = T_{ant\_noise} + T_{LNA} + T_{cpl\_simul} \quad (1)$$

Table I compares the result of (1) with the measured array noise temperature $T_{total\_measm}$, from which a 'measured' value for the noise coupling may be derived, according to the following formula:

$$T_{cpl\_measm} = T_{total\_measm} - T_{ant\_noise} - T_{LNA} \quad (2)$$

**Table I  Noise budget in [K] of the 49-element array as a function of scan angle at 1400 MHz for the north-south direction**

| Scan angle (°) from broadside | 0 | 10 | 20 | 30 | 40 | 50 |
|---|---|---|---|---|---|---|
| $T_{ant\_noise}$ | 4 | 4 | 4 | 4 | 4 | 4 |
| $T_{LNA}$ | 38 | 38 | 38 | 38 | 38 | 38 |
| $T_{cpl\_measm}$ | 7 | 7 | 11 | 9 | 15 | 24 |
| $T_{total\_measm}$ | 49 | 49 | 53 | 51 | 57 | 66 |
| $T_{cpl\_simul}$ | 6 | 6 | 7 | 9 | 11 | 18 |
| $T_{total\_simul}$ | 48 | 48 | 49 | 51 | 53 | 60 |

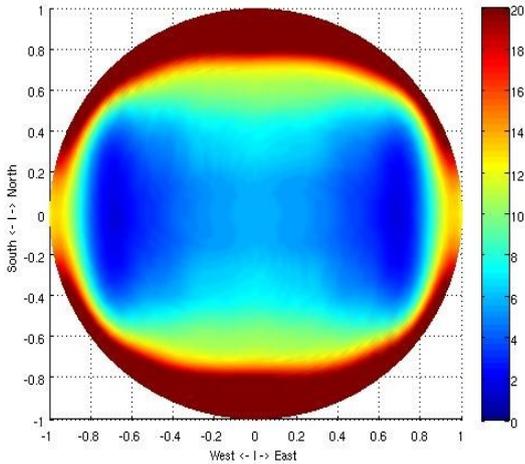

**Fig. 6 Two-dimensional scan of the simulated noise coupling contribution in [K] for the 49-element array at 1400 MHz**

The simulated noise coupling contributions at 1400 MHz for a full two-dimensional scan are presented in Fig. 6 for one polarization of the array, showing relatively low noise coupling over a large scan range. One-dimensional scans in the direction with the smallest change in noise coupling (east-west), as well as for the perpendicular direction are shown in Fig 7. The noise coupling contribution remains below 10 K for scan angles up to 60º in the east-west direction and below 14 K over the full scan range. In the north-south direction a maximum value of 13 K for the noise coupling at 1400 MHz was calculated at 45º, increasing to 50 K for scanning near the horizon. This result underlines that the increase in measured system noise temperature at large scan angles in Fig. 5 is caused by noise pick up at the horizon.

The sum of the simulated noise contributions and the LNA noise temperature in Table I is almost identical (within the measurement accuracy of 5 K) to the measured array noise temperature for small scan angles. This confirms the validity of the models, with the active reflection coefficient being the only variable as a function of scan angle in Figs. 6 and 7 and for small scan angles in Fig. 5. For scan angles larger than 40-50° the measured results in Figs. 4 and 5 are influenced by noise pick up from the environment. The presented results lead to the prediction that for larger arrays with a narrow beam (and relatively low side-lobes) the simulation results will more accurately predict the measurements at larger scan angles. This will be the subject of further study.

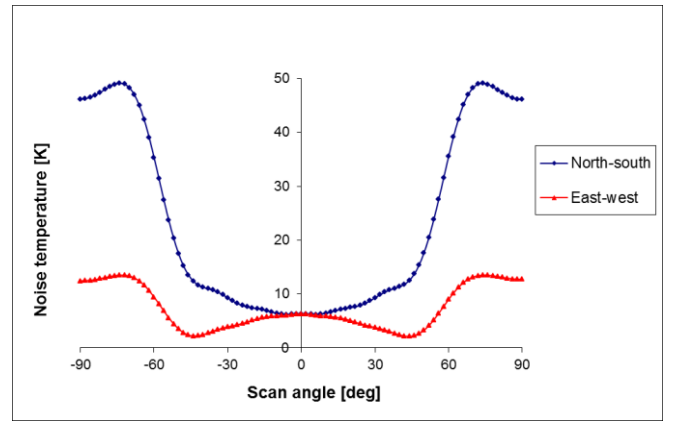

**Fig. 7 Noise coupling contribution at 1400 MHz for a 49-element array, for two orthogonal scan directions**

V. CONCLUSIONS

In this paper experimental noise temperature results for an AA-tile have been presented, which demonstrate the lowest array receiver noise temperature for an AA-tile to date, with a minimum measured value of 45 K in L-band. The properties as a function of frequency and the effects of scanning the beam on the noise temperature are shown and compared to simulation results, showing good agreement. Based on the measurement and simulation results a number of observations have been made, which lead to the following conclusions:
- array noise temperatures below 50 K have been consistently measured, with a value of 45 K at 1200 MHz
- the use of the edge elements in the array may cause some increase in system noise temperature, but lead to lower noise temperature if a smaller part of the array is active
- for scan angles up to ~45º the increase in noise temperature due to noise coupling remains below 13 K
- a beam formed with more elements may be scanned to larger angles before the horizon introduces noise, due to the narrow beam
- scanning a smaller beam to the horizon results in a larger contribution from the horizon to the system noise, because the beam then 'sees' a larger part of a hotter environment

- verification that the variation in system noise temperature for large scan angles is caused solely by a change in the noise coupling contribution, can only be done with a larger array, i.e. with a narrow beam (with low side-lobes)

The presented results illustrate the low-noise potential of AAs for application as future sensitive radio telescopes.